\documentclass[published]{JHEP3} % 10pt is ignored!

\JHEP{00(2010)000}

\JHEPspecialurl{http://jhep.sissa.it/JOURNAL/JHEP3.tar.gz}

\usepackage{epsfig,multicol,bbm}
\usepackage{amsmath}
\usepackage{amsfonts}
\usepackage{amssymb}
\def\c #1{{\cal #1}}                             % calligraphic letter
\def\Dirac{{\raise0.09em\hbox{/}}\kern-0.69em D}% Dirac operator
\def\ep{i\epsilon} 

\def\kbar{{\mathchar'26\mkern-9muk}} 		% kbar symbol
\def\lesssim{\mathrel{\hbox{\rlap{\hbox{\lower4pt\hbox{$\sim$}}}\hbox{$<$}}}}
\def\sq{\hbox{\rlap{$\sqcap$}$\sqcup$}}         % d'Alembertian
\def\p{\partial}                                %\partial derivative 
\def\beg{\begin{eg}\rm}                         % Macro to begin an Example
\def\eeg{\hfill\sq\end{eg}}                     % Macro to end an Example

\def\k {\kern-.1em\mathbin{,}\kern-.1em}
\def\hk{\kern.12em\raise-1em\hbox{$\hat{\raise1em\hbox{,}}$}\kern.12em}

\newcounter{eg}                                 %\newcounter{chapter} 
\newtheorem{eg}{Example}[section]
\def\beg{\begin{eg}\rm}                         %Macro to begin an Example
\def\eeg{\hfill\sq\end{eg}}                     % Macro to end an Example
        
\newcommand{\initiate}{\setcounter{equation}{0}}        
\newcommand\fverb{\setbox\fverbbox=\hbox\bgroup\verb}
\newcommand\fverbdo{\egroup\medskip\noindent%
			\fbox{\unhbox\fverbbox}\ }
\newcommand\fverbit{\egroup\item[\fbox{\unhbox\fverbbox}]}
\newbox\fverbbox

\title{Geometry of the Grosse-Wulkenhaar Model}

\author{Maja Buri\'c $^{1}$ and Michael Wohlgenannt $^{2}$
\\\vskip 10pt
$\strut^{1}$Faculty of Physics \\
University of Belgrade, P.O. Box 368\\
SR-11001 Belgrade, Serbia\\
Email: majab@phy.bg.ac.yu
\\
\vskip 3pt
$\strut^{2}$Institute for Theoretical Physics\\
Vienna University of Technology\\
Wiedner Hauptstr. 8-10\\
A-1040 Vienna, Austria\\
Email: michael.wohlgenannt@tuwien.ac.at}

\received{\today} 		%%
\accepted{\today}		%% These are for published papers.

\preprint{\hepth{0902.3408}}	% OR: \preprint{Aaaa/Mm/Yy\\Aaa-aa/Nnnnnn}
\abstract{We analyze properties of a family of finite-matrix spaces obtained
by a truncation of the Heisenberg algebra and we show that it has a
three-dimensional, noncommutative and curved geometry. Further, we demonstrate
that the Heisenberg algebra can be described as a two-dimensional hyperplane 
embedded in this space. As a consequence of the given construction we show 
that the Grosse-Wulkenhaar (renormalizable) action can be interpreted as 
the action for the scalar field on a curved background space. We discuss 
the generalization to four dimensions.}

\keywords{noncommutative geometry, renormalization, noncommutative $\phi^4$ model}
\begin{document}

\section{Motivation and introduction}

Of all noncommutative spaces the Heisenberg algebra, that is the space with constant noncommutativity of coordinates,
\begin{equation}
 [x,y] = i\kbar                                \label{heis}
\end{equation} 
has a special role, mainly due to our century-long experience with quantum mechanics. 
In particular, field theories on (\ref{heis}) have been defined in various versions
and their classical and quantum properties were analyzed in details. Fields on the 
Heisenberg algebra are usually represented by functions on {\bf R}$^n$  with multiplication  
given by  the Moyal-Weyl product. This representation is mathematically well understood and intuitively appealing; moreover it has an apparent commutative limit $\kbar \to 0$. Field theories on other algebras, for example on the fuzzy sphere or on the fuzzy $CP^n$, have also been discussed but certainly not so extensively, \cite{cp}. 

The main advantage of a field theory defined on a space having the structure of 
Lie algebra with finite-dimensional representations is its finiteness upon quantization:
The integral is a trace of a matrix and the functional integration reduces to a well-defined finite expression, namely to an integral over the finite-dimensional space of matrices. This is a 
feature which one intuitively expects from a theory on a noncommutative space:
to regularize divergences. The problem with the Lie-algebra spaces 
is usually in the definition of a relevant commutative limit, especially if that limit
is the flat Minkowski space. 

Renormalizability of field theories on the Heisenberg algebra on the other hand is a long discussed issue. If we focus on the scalar field theory defined by action
\begin{equation}
 S = \int  \frac 12\, \p_\mu \varphi\, \p^\mu \varphi +\frac {m^2}{2 }\,\varphi^2 +\frac{\lambda}{4!}\,\varphi^4 ,                                               \label{free}
\end{equation} 
 the calculations done so far converge to the conclusion that the theory is not renormalizable, at least within the usual perturbative schemes, \cite{chep}. For a different approach, see \cite{Cho:1999sg}. An important exception which was singled out in the last years was found by H.~Grosse and R.~Wulkenhaar \cite{Grosse:2003nw,Grosse:2004yu}: If we add to the action (\ref{free}) the harmonic oscillator potential term
\begin{equation}
 S^\prime =  \int  \frac 12\, \p_\mu \,\varphi \p^\mu \varphi +\frac {m^2}{2 }\,\varphi^2 +\frac{\Omega^2}{2} \,{\tilde x}^\mu \varphi\, {\tilde x}_\mu \varphi +\frac{\lambda}{4!}\,\varphi^4 , 
      \label{GW4}
\end{equation} 
the corresponding theory is renormalizable. 
The physical reason behind this is the additional symmetry which (\ref{GW4}) possesses called 
the Langmann-Szabo duality, \cite{LS}.  This symmetry interchanges the UV and IR sectors of the theory.  We shall show that the action (\ref{GW4}) has another interesting property, namely that the oscillator term can be interpreted geometrically: It is the coupling of the scalar field to the curvature of an appropriately defined noncommutative space. 

The plan of the paper is the following:
In the initial sections we define and study in some detail the geometry of 
the  truncted Heisenberg space, whose commutation
relations are a specific combination of a quantum group and a Lie algebra.
We then  analyze the properties of its subspace $z=0$. We  show in 
Section 5 that from the point of view of the algebra and its representations
the hyperplane $z=0$ is a two-dimensional subspace of the
truncated Heisenberg algebra. But the dimensionality of the
corresponding space of 1-forms  is three,
and moreover the hyperplane is not flat but curved. Finally, in Section 6 we show
that the action for the scalar field coupled to the curvature of the described
background space equals the action of the Grosse-Wulkenhaar model.
Some additional remarks concerning the properties of the geometry defined by the
noncommutative frame formalism are given in the appendices.

\section{Notation and basic formalism}

Let us introduce the notation. Suppose that the noncommutative space, that is the algebra $\c{A}$, is generated by hermitian elements, linear operators $x^\mu$ ($\mu = 1,\dots,n$) 
which satisfy the commutation relation
\begin{equation}
 [x^\mu,x^\nu] = \ep J^{\mu\nu}(x) .                  \label{a}
\end{equation} 
In the special case of Heisenberg algebra (\ref{heis}) we have 
$J^{\mu\nu}=J^{12}\,$=\,const. Along with coordinate indices $\mu,\, \nu$ we shall 
use frame indices $\alpha,\, \beta$  ($\, = 1,\dots,n$) to denote the components 
of vectors and 1-forms in the moving frame basis. The constant $\epsilon$ 
which is introduced in 
(\ref{a}) is a parameter which, similarly to $\kbar$ in (\ref{heis}), measures 
noncommutativity. We shall assume that $\epsilon$ is dimensionless; the commutative limit is given by  $\epsilon\to 0$.

If $J^{\mu\nu}$\,=\,const the noncommutative space can be endowed with a flat connection 
and a flat metric. The differential $d$ which corresponds to this choice is defined by 
imposing the following commutation relations on the algebra of 1-forms
\begin{eqnarray}
 &&[x^\mu, dx^\nu] = 0,                                              \label{b}\\[4pt]
&&
 [dx^\mu, dx^\nu] = dx^\mu dx^\nu  + dx^\nu dx^\mu = 0     ,            \label{c}
\end{eqnarray} 
and the Leibniz rule. Obviously, differential calculus (\ref{b}-\ref{c}) is consistent 
with the initial algebra.  In fact relation  (\ref{b}) shows that 1-forms $dx^\mu$ can be identified with the elements $\theta^\alpha$ of a noncommutative moving frame (vielbein)
\begin{equation}
 \theta^\alpha = \delta^\alpha_\mu dx^\mu         ,                      \label{frameH}
\end{equation} 
because property $ [x^\mu, \theta^\alpha] = 0 $
is sufficient to insure that the frame components of the metric  are constant, which
means that the space is locally flat\footnote{The requirement $g^{\alpha\beta}$ = const, 
that is $[x^\mu,g^{\alpha\beta}]=0\,$, is a very stringent one. Here it is, by linearity of the metric, transferred to a less constraining condition $\, [x^\mu, \theta^\alpha] = 0 $.}. 
Derivations $\p_\mu f$, defined by relation
\begin{equation}
 df = \p_\mu f\, dx^\mu = e_\alpha f\, \theta^\alpha ,
\end{equation} 
are in this case inner, $ \p_\mu f =[p_\mu,f] $, and
generated by momenta $p_\mu$,
\begin{equation}
 p_\mu = \frac{1}{\ep}(J^{-1})_{\mu\nu} x^\nu ,             \label{P}
\end{equation} 
assuming of course that the matrix $J^{\mu\nu}$ is not degenerate.
If we  calculate the Ricci rotation coefficients $ C^\alpha{}_{\beta\gamma}$ from 
the definition
\begin{equation}
 d\theta^\alpha = -\frac 12 C^\alpha{}_{\beta\gamma} \theta^\beta\theta^\gamma,   \label{C}
\end{equation}  
we see that $C^\alpha{}_{\beta\gamma} $ vanish for (\ref{frameH}), that is the space 
is also globally flat. As one can easily see,  momenta $p_\alpha$ can be used to generate 
the algebra $\c{A}$
instead of  coordinates $x^\mu$; relation (\ref{P}) is a kind of Fourier transformation. 
The momenta satisfy in the Heisenberg case a commutation relation  similar to (\ref{heis}):
\begin{equation}
 [p_\mu,p_\nu] = - \frac{1}{\ep} (J^{-1})_{\mu\nu} ={\rm const}.
\end{equation}

It is in principle possible to define differential structures for arbitrary dependence of the commutator $J^{\mu\nu}(x)$. One has to find a set of 1-forms $\theta^\alpha$ 
which commute with all elements of $\c{A}$ consistently with commutation relations
(\ref{a});  this set is then a  frame. The choice of the frame might not be unique: the same noncommutative `manifold' can support, in principle, different noncommutative geometries.
In the special case of the matrix spaces all derivations are necessarily inner: the definition 
of the differential $d$ reduces then to the choice of $ p_\alpha$. However the set of
$ p_\alpha $ is not completely arbitrary. It can be shown that, in order to 
have relations $\, d\, [x^\mu,x^\nu] = \ep \, dJ^{\mu\nu}\,$ and $d^2=0\,$ fulfilled, 
the momenta have to satisfy a quadratic algebra, \cite{book}:
\begin{equation}
 [p_\alpha,p_\beta] = \frac{1}{\ep}K_{\alpha\beta} +F^\gamma{}_{\alpha\beta}p_\gamma -2\ep Q^{\gamma\delta}{}_{\alpha\beta}p_\gamma p_\delta ,                 \label{pp}
\end{equation} 
 where $ K_{\alpha\beta}$, $F^\gamma{}_{\alpha\beta} $ and $ Q^{\gamma\delta}{}_{\alpha\beta}$ are constants or belong to the center of $\c{A}$.
This requirement is an elementary consistency constraint on  possible differential structures.
As we shall see there are other constraints which make the choice of the differential almost
unique.

\initiate
\section{The truncated Heisenberg algebra}

Heisenberg algebra (\ref{heis}) can  be represented in the Fock basis (that is,
 in the energy representation of the harmonic oscillator) by infinite-dimensional 
matrices as
\begin{eqnarray}
&& x = \frac{1}{\sqrt{2}}
\begin{pmatrix}
0 & 1 &0 & .&. &.& . \cr
1 &0 &\sqrt{2}&. & .&. &.\cr
0  &\sqrt{2} & 0&. &. &. &.\cr
. &. &.  &.& .& . & . \cr
. & . &. &.& 0 & \sqrt{n-1} & .\cr
. & . &.& .&\sqrt{n-1}& 0 & .\cr
. &. &. &.  & .&. & .
\end{pmatrix},                                        \label{x}
\\[10pt]
&&y = \frac{i}{\sqrt{2}}
\begin{pmatrix}
0 & -1 &0 & . &.&.& . \cr
1 &0 &-\sqrt{2} &.& .&. &.\cr
0  &\sqrt{2} & 0&. &. &.&.\cr
. &. &.  &.& . &.& . \cr
. & . &.&. & 0 & -\sqrt{n-1} & .\cr
. & . &.& .&\sqrt{n-1}& 0 & .\cr
. &. &. &.&.  & . & .
\end{pmatrix}.                                     \label{y}
\end{eqnarray} 
 As it is usual in  quantum mechanics, $x$ and $y$ are in (\ref{x}-\ref{y})
taken to be dimensionless.
Truncation from $\infty\times\infty$ to  $n\times n$ matrices consisting of 
the first $n$ rows and the first $n$ columns,
\begin{equation}
x_n = \frac{1}{\sqrt{2}}
\begin{pmatrix}
0 & 1 &0 &.& . &. \cr
1 &0 &\sqrt{2}& .&. &. \cr
0  &\sqrt{2} & 0&. &.& .\cr
. &. &.  &. & . &. \cr
. & . & .&. &0 & \sqrt{n-1} \cr
. & . & .& .&\sqrt{n-1}& 0 
\end{pmatrix},
\end{equation} 

\begin{equation}
y_n = \frac{i}{\sqrt{2}}
\begin{pmatrix}
0 & -1 &0 & . &.&. \cr
1 &0 &-\sqrt{2}& .&. &. \cr
0  &\sqrt{2} & 0&. &.&.\cr
. &. &.  &. & . &. \cr
. & . & . &.&0 & -\sqrt{n-1} \cr
. & . & .& .&\sqrt{n-1}& 0 
\end{pmatrix} ,
\end{equation} 
changes the initial algebra $\, [x,y] = i \ $   to 
\begin{equation}
 [x_n,y_n] = i(1-nP_n),                                   \label{trheis}
\end{equation} 
\noindent
where $P_n$ denotes the projector
\begin{equation}
 P_n =  
\begin{pmatrix} 
0 & 0 & 0&. & .& . \cr
0 & 0 & 0&. & .&. \cr
0 & 0 & 0&. & .&. \cr
. & . & .&. & .&. \cr
. & . &.& .& 0 &0\cr
. & .& .&. & 0 & 1  
\end{pmatrix}  .
\end{equation}
The limit $n\to\infty$ in which (\ref{trheis}) becomes the Heisenberg algebra is a weak 
limit; note that it can be written  formally as $P_n = 0$, or  $n P_n = 0$ as well.

Matrices $x_n$ and $y_n$ have nice geometric interpretation: for fixed $n$, 
they describe a finite part of the two-dimensional plane. One can see this from
the spectrum of $x_n$, $y_n$ which consists of all zeroes of the Hermite 
polynomials $H_n$, \cite{Polychronakos:2007df}, and therefore the expectation 
values of $x_n$ and $ y_n$ are bounded by the largest zero of $H_n$.
 When $n$ grows, $x_n$ and $y_n$  approximate larger and larger part of the plane 
with more and more points, of course not densely. In the limit $n \to\infty$ one 
obtains the whole noncommutative $x$--$y$  plane.

We designate algebra (\ref{trheis}) the `truncated Heisenberg algebra' because 
it is obtained by truncation from infinite to finite matrices. In the following 
we will omit the index $n$ as we shall write it in the form without an explicit 
$n$-dependence. We define as usual
\begin{equation}
a =  \frac{1}{\sqrt{2}} (x+iy) = 
\begin{pmatrix}
0 & 1 &0 & . &.&. \cr
0 &0 &\sqrt{2}& .&. &. \cr
0  &0 & 0&. &.& .\cr
. &. &.  &. & . &. \cr
. & . & . &.&0 & \sqrt{n-1} \cr
. & . & .& .&0& 0 
\end{pmatrix}
 ,\quad N =a^\dagger a ,
\end{equation} 
 and so on.
For fixed $n$ there are additional relations in the algebra, for example
\begin{equation}
 a^n =0,\qquad Pa =0,\qquad a^{n-1}(1-P) =0 ,\qquad P^2=P.        \label{**}
\end{equation} 

Another well known  finite matrix approximation of the Heisenberg algebra 
was given by Holstein and Primakoff, \cite{Holstein:1940zp}. 
For that approximation too the Heisenberg algebra is obtained in the formal limit 
$n\to\infty$. From the eigenvalues of coordinate operators one sees that the two 
given approximations are not unitarily equivalent. The geometry of the
 Holstein-Primakoff space, given by the differential calculus defined in \cite{book}, 
is also different: it is two-dimensional and flat. 
 
The truncated Heisenberg algebra can be viewed as a three-dimensional noncommutative 
space generated by coordinates $x$, $y$ and $P$. The commutation relation 
\begin{equation}
 [x,y] =i (1-nP)                                  \label{1}
\end{equation} 
then has to be completed with the two missing relations.
From  $Pa=0$ and its adjoint $a^\dagger P =0$ we have
\begin{equation}
 [x,P] = i(yP+Py)  ,   \qquad                                   
 [y,P] = - i(xP+Px)  .                                       \label{3}
\end{equation} 
In order to think more abstractly we denote $nP=z$ and write the algebra in the form
\begin{eqnarray}
 &&[x,y] =i (1-z) ,                                            \label{1z}\\[4pt]
&& [x,z] = i(yz+zy)  ,                                       \nonumber\\[4pt]
&& [y,z] = - i(xz+zx)  .                                       \nonumber
\end{eqnarray} 
The remaining relations from (\ref{**}) 
 need not to be included in the algebra (\ref{1z}), or more precisely
in its momentum version (\ref{ppp}): what is important is that they are stable
under differentiation. However the last relation in (\ref{**})
 can be used write  the algebra in another form: We shall return
to this point in more detail in Appendix 1\footnote{We thank the referee for
raising this interesting question.}.
One should  keep in mind that the truncated Heisenberg algebra has finite-dimensional representations for 
all $n$. As we shall see, the formal limit $n\to\infty$ can be consistently viewed as 
an embedding of the hyperplane $z=0$ in the given space.
Algebra (\ref{1z})  is quadratic in its generators: This will allow us to identify the 
momenta easily and to define a differential calculus. 

Before proceeding to that, let us introduce physical dimensions in (\ref{1z}).
The parameter $\epsilon$ in (\ref{a})  is dimensionless and defined in such a way 
that $\epsilon\to 0$ gives the commutative limit. In fact we have at least two relevant 
length scales in the problem. One of them is $\sqrt\kbar$, the scale at which effects of noncommutativity become important. The other scale, as we are presumably dealing with gravity, 
is the gravitational scale which we denote by $\mu^{-1}$ (the Schwarzschild radius or the cosmological constant for example). We assume that $\epsilon = \mu^2\kbar$. 
Therefore we write the algebra as
\begin{eqnarray}
 &&[x,y] = i \epsilon \mu^{-2}(1-\mu z) ,                                            \label{zz}\\[4pt]
&& [x,z]  = i\epsilon(yz+zy) ,                                      \nonumber\\[4pt]
&& [y,z] = - i\epsilon(xz+zx)            .                           \nonumber
\end{eqnarray}

\initiate
\section{Differential geometry in the frame formalism}

Noncommutative differential geometry which we use is defined by a generalization
of the moving frame formalism of Cartan. We will briefly review some of its 
properties; a more detailed exposition can be found in \cite{book}. In the case when
momenta  $ p_\alpha$ generate the space one can express all quantities in terms of
them instead as functions of coordinates. This simplifies calculations because momenta 
obey quadratic relation of a fixed form. The latter can also be written as
\begin{equation}
 2P^{\gamma\delta}{}_{\alpha\beta}p_\gamma p_\delta -F^\gamma{}_{\alpha\beta} p_\gamma -\frac{1}{\ep} K_{\alpha\beta} =0 .                  \label{2*}
\end{equation} 
Constants $ P^{\gamma\delta}{}_{\alpha\beta}$ define exterior multiplication of 1-forms:
\begin{equation}
 \theta^\gamma\theta^\delta = P^{\gamma\delta}{}_{\alpha\beta}\theta^\alpha\theta^\beta,
                                         \label{pro}
\end{equation} 
and thus $ P^{\gamma\delta}{}_{\alpha\beta}$ is a projection. In the commutative case $P^{\gamma\delta}{}_{\alpha\beta} $ is the antisymmetrization,
\begin{equation}
  P^{\gamma\delta}{}_{\alpha\beta} = \frac 12 (\delta^\gamma_\alpha \delta^\delta_\beta -
\delta^\gamma_\beta \delta^\delta_\alpha) =: \frac 12 \delta^{\gamma\delta}_{\alpha\beta},
\end{equation} 
while in the noncommutative case we can write 
\begin{equation}
  P^{\gamma\delta}{}_{\alpha\beta} =  \frac 12 \delta^{\gamma\delta}_{\alpha\beta}+\ep Q^{\gamma\delta}{}_{\alpha\beta}.                         \label{*}
\end{equation}

Requirements on hermiticity of the frame forms, hermiticity of the exterior product etc. 
give additional constraints which we will not discuss here, see \cite{reality}. We will however use the fact 
that coefficients $Q^{\gamma\delta}{}_{\alpha\beta}$ are symmetric in the upper and 
antisymmetric in the lower pair of indices, evident from (\ref{1z}). From the general 
formalism one can show that the Ricci rotation coefficients (\ref{C}) are linear in momenta 
and equal to
\begin{equation}
 C^\gamma{}_{\alpha\beta} = F^\gamma{}_{\alpha\beta} -4\ep Q^{\gamma\delta}{}_{\alpha\beta}p_\delta.
\end{equation} 

Differential-geometric quantities are defined in complete analogy with the commutative case,
which includes for example the requirement of linearity. The (inverse) metric in the frame basis has constant components
\begin{equation}
 g^{\alpha\beta} = g(\theta^\alpha\otimes\theta^\beta) = {\rm const}.
\end{equation} 
The connection
$\omega^\alpha{}_\beta =\omega^\alpha{}_{\gamma\beta} \theta^\gamma$
and the torsion 
$\Theta^\alpha = \Theta^\alpha{}_{\gamma\beta}\theta^\gamma\theta^\beta$ 
 are related  by the structure equation $
 \Theta^\alpha = d\theta^\alpha +\omega^\alpha{}_\beta\theta^\beta $.
One can impose additional relations among $g$, $\omega^\alpha{}_\beta$ and $\Theta^\alpha$:
To formulate them it is necessary to introduce a mapping which reverses the order of 
indices in the tensor product of 1-forms -- the `flip' $\sigma$:
\begin{equation}
 \sigma(\theta^\gamma\otimes\theta^\delta) = 
S^{\gamma\delta}{}_{\alpha\beta}\theta^\alpha\otimes\theta^\beta .           \label{flip}
\end{equation} 
Coefficients $S^{\gamma\delta}{}_{\alpha\beta} $ are constants and  reduce in the
commutative limit to $ \delta^\gamma_\beta\delta^\delta_\alpha$; we write them as
\begin{equation}
 S^{\gamma\delta}{}_{\alpha\beta} = \delta^\gamma_\beta\delta^\delta_\alpha +\ep T^{\gamma\delta}{}_{\alpha\beta} .
\end{equation} 

It  seems natural to require that the connection be metric-compatible and that the torsion vanish, as these conditions are rather usual and can  always  be imposed in commutative geometry. 
Here they are expressed as
\begin{eqnarray}
&& \omega^\alpha{}_{\beta\gamma} g^{\gamma\delta} +\omega^{\delta}{}_{\gamma\epsilon}S^{\alpha\gamma}{}_{\beta\eta} g^{\eta\epsilon} =0,
                                        \label{MC}
\\[4pt]
&& \omega^\alpha{}_{\beta\gamma} P^{\beta\gamma}{}_{\delta\epsilon} =\frac 12 C^\alpha{}_{\delta\epsilon}   ,                \label{TF}
\end{eqnarray} 
respectively. However, (\ref{MC}-\ref{TF}) is a set of nonlinear algebraic relations in 
coefficients $g^{\alpha\beta}$, $T^{\alpha\beta}{}_{\gamma\delta}$, 
$F^{\alpha}{}_{\beta\gamma}$ and $Q^{\alpha\beta}{}_{\gamma\delta}$ and it is not 
obvious that solutions exist in nontrivial cases. But one can impose these conditions 
in the commutative limit: for $\epsilon\to 0$, equations (\ref{MC}-\ref{TF}) can be 
linearized and solved, \cite{book}. The solution has the same form as in  commutative geometry:
\begin{equation}
 \omega_{\alpha\beta\gamma} =\frac 12 ( C_{\alpha\beta\gamma} - C_{\beta\gamma\alpha} +C_{\gamma\alpha\beta})    ,                \label{omega}
\end{equation} 
or equivalently
\begin{equation}
 \omega^\alpha{}_{\beta\gamma} =\frac 12 \, F^\alpha{}_{\beta\gamma} +\ep T^{\alpha\delta}{}_{\beta\gamma} p_\delta  ,                     \label{connection}
\end{equation} 
with
\begin{equation}
 F^{(\alpha\beta\gamma)} =0, \qquad T^{(\alpha\delta}{}_\beta{}^{\gamma)} =0.
\end{equation} 
From (\ref{omega}) we obtain the expression for the coefficients 
$ T_{\alpha\beta\gamma\delta} $:
\begin{equation}
 T_{\alpha\beta\gamma\delta} = 2( -Q_{\alpha\beta\gamma\delta} +Q_{\beta\gamma\delta\alpha} +Q_{\beta\delta\gamma\alpha}) ;                         \label{T}
\end{equation} 
in particular,
\begin{equation}
 T^{\alpha\beta}{}_{[\gamma\delta]} =-4 Q^{\alpha\beta}{}_{\gamma\delta} .
\end{equation} 
These relations will be used in the calculation of the curvature.

The Riemann curvature is defined by the usual formula
\begin{equation}
 \Omega^\alpha{}_\beta =d\omega^\alpha{}_\beta +\omega^\alpha{}_\gamma\omega^\gamma{}_\beta = \frac 12 \, R^\alpha{}_{\beta\rho\sigma}\theta^\rho\theta^\sigma . 
\end{equation}  
Calculating its coefficients in terms of the momenta we obtain
\begin{eqnarray}
 &&R^\alpha{}_{\beta\rho\sigma}\theta^\rho\theta^\sigma= 2 \Big( T^{\alpha\gamma}{}_{\sigma\beta} K_{\rho\gamma} -\frac 14 F^\alpha{}_{\gamma\beta}F^\gamma{}_{\rho\sigma} + \frac 14 F^\alpha{}_{\rho\gamma} F^\gamma{}_{\sigma\beta} \\[6pt]
&&\phantom{R}+\ep p_\zeta( F^\zeta{}_{\rho\gamma}T^{\alpha\gamma}{}_{\sigma\beta} + F^\alpha{}_{\gamma\beta}Q^{\gamma\zeta}{}_{\rho\sigma} -\frac 12 F^\gamma{}_{\rho\sigma}T^{\alpha\zeta}{}_{\gamma\beta}+\frac 12 F^\alpha{}_{\rho\gamma}T^{\gamma\zeta}{}_{\sigma\beta} +\frac 12 F^\gamma{}_{\sigma\beta}T^{\alpha\zeta}{}_{\rho\gamma}
)  \nonumber \\[6pt]
&&\phantom{R}+(\ep)^2 p_\zeta p_\eta (-2T^{\alpha\gamma}{}_{\sigma\beta}Q^{\zeta\eta}{}_{\rho\gamma}+2T^{\alpha\zeta}{}_{\gamma\beta}Q^{\gamma\eta}{}_{\rho\sigma}+T^{\alpha\zeta}{}_{\rho\gamma}T^{\gamma\eta}{}_{\sigma\beta}
)\Big) \theta^\rho\theta^\sigma ,\nonumber
\end{eqnarray} 
that is
\begin{eqnarray}
 &&R^\alpha{}_{\beta\delta\epsilon}= 2 \Big( T^{\alpha\gamma}{}_{\sigma\beta} K_{\rho\gamma} -\frac 14 F^\alpha{}_{\gamma\beta}F^\gamma{}_{\rho\sigma} + \frac 14 F^\alpha{}_{\rho\gamma} F^\gamma{}_{\sigma\beta} \label{rr} \\[6pt]
&&\phantom{R}+\ep p_\zeta( F^\zeta{}_{\rho\gamma}T^{\alpha\gamma}{}_{\sigma\beta} + F^\alpha{}_{\gamma\beta}Q^{\gamma\zeta}{}_{\rho\sigma} -\frac 12 F^\gamma{}_{\rho\sigma}T^{\alpha\zeta}{}_{\gamma\beta}+\frac 12 F^\alpha{}_{\rho\gamma}T^{\gamma\zeta}{}_{\sigma\beta} +\frac 12 F^\gamma{}_{\sigma\beta}T^{\alpha\zeta}{}_{\rho\gamma}
)  \nonumber \\[6pt]
&&\phantom{R}+(\ep)^2 p_\zeta p_\eta (-2T^{\alpha\gamma}{}_{\sigma\beta}Q^{\zeta\eta}{}_{\rho\gamma}+2T^{\alpha\zeta}{}_{\gamma\beta}Q^{\gamma\eta}{}_{\rho\sigma}+T^{\alpha\zeta}{}_{\rho\gamma}T^{\gamma\eta}{}_{\sigma\beta}
)\Big) P^{\rho\sigma}{}_{\delta\epsilon} . \nonumber
\end{eqnarray} 
The curvature is a second-order polynomial in the momenta. Note that, as momenta are defined by a relation of the type (\ref{P}), $\ep p_\alpha$ is of the same order of magnitude as  $x^\mu$ even if $\epsilon$ is small. We write the curvature  as the sum of two terms  
\begin{equation}
 R^\alpha{}_{\beta\epsilon\delta}= {R_0}^\alpha{}_{\beta\epsilon\delta} +\ep{R_1}^\alpha{}_{\beta\epsilon\delta} ,
\end{equation} 
with
\begin{eqnarray}
 &&{R_0}^\alpha{}_{\beta\epsilon\delta}=  \Big( T^{\alpha\gamma}{}_{\sigma\beta} K_{\rho\gamma} + \frac 14 F^\alpha{}_{\rho\gamma} F^\gamma{}_{\sigma\beta}         \label{r0}    \\[6pt]
&&\phantom{R}+\ep p_\zeta( F^\zeta{}_{\rho\gamma}T^{\alpha\gamma}{}_{\sigma\beta} +\frac 12 F^\alpha{}_{\rho\gamma}T^{\gamma\zeta}{}_{\sigma\beta} +\frac 12 F^\gamma{}_{\sigma\beta}T^{\alpha\zeta}{}_{\rho\gamma}
)  \nonumber \\[6pt]
&&\phantom{R}+(\ep)^2 p_\zeta p_\eta (-2T^{\alpha\gamma}{}_{\sigma\beta}Q^{\zeta\eta}{}_{\rho\gamma}+T^{\alpha\zeta}{}_{\rho\gamma}T^{\gamma\eta}{}_{\sigma\beta}
)\Big) \delta^{\rho\sigma}_{\epsilon\delta}                            \nonumber
\end{eqnarray} 
and
\begin{eqnarray}
 &&{R_1}^\alpha{}_{\beta\epsilon\delta}= 2 \Big( T^{\alpha\gamma}{}_{\sigma\beta} K_{\rho\gamma} + \frac 14 F^\alpha{}_{\rho\gamma} F^\gamma{}_{\sigma\beta}         \label{r1}          \\[6pt]
&&\phantom{R}+\ep p_\zeta( F^\zeta{}_{\rho\gamma}T^{\alpha\gamma}{}_{\sigma\beta} +\frac 12 F^\alpha{}_{\rho\gamma}T^{\gamma\zeta}{}_{\sigma\beta} +\frac 12 F^\gamma{}_{\sigma\beta}T^{\alpha\zeta}{}_{\rho\gamma}
)  \nonumber \\[6pt]
&&\phantom{R}+(\ep)^2 p_\zeta p_\eta (-2T^{\alpha\gamma}{}_{\sigma\beta}Q^{\zeta\eta}{}_{\rho\gamma}+T^{\alpha\zeta}{}_{\rho\gamma}T^{\gamma\eta}{}_{\sigma\beta}
)\Big) Q^{\rho\sigma}{}_{\epsilon\delta}  .\nonumber
\end{eqnarray} 
Contracting the Riemann curvature from~(\ref{r0}-\ref{r1}) we obtain for
 the curvature scalar $R = g^{\beta\delta} R^\alpha{}_{\beta\alpha\delta} = R_0+\ep R_1\,$ the following expression:
\begin{eqnarray}
 && R_0= - 8 K_{\alpha\gamma} Q^{\alpha\beta}{}_\beta{}^\gamma + \frac 14 F^{\alpha\beta\gamma} F_{\alpha\beta\gamma}                          \label{R}
+2 \ep p_\zeta (- 4 F^\zeta{}_{\alpha\gamma} Q^{\alpha\beta}{}_\beta{}^\gamma  - F_{\alpha\beta\gamma} Q^{\zeta\alpha\beta\gamma} )
 \\[6pt] 
&& \phantom{R}
+ 4 (\ep)^2 p_\zeta p_\eta ( 4 Q^{\alpha\beta}{}_\beta{}^\gamma Q^{\zeta\eta\alpha\gamma} - 4 Q^{\gamma\beta}{}_\beta{}^\zeta Q_{\gamma\alpha}{}^{\alpha\eta} 
-Q^{\zeta\alpha\beta\gamma} Q^\eta{}_{\alpha\beta\gamma} +2 Q^{\zeta\alpha\beta\gamma} Q^\eta{}_{\gamma\alpha\beta} ) , \nonumber\\[16pt]
&&
  R_1= 4(\ep) [p_\rho,p_\gamma]\, T^{\alpha\gamma}{}_\sigma{}^\beta Q^{\rho\sigma}{}_{\alpha\beta}
- (\ep)^2 [p_\zeta,p_\eta]\, T^{\gamma\zeta\rho\alpha} T_\gamma{}^{\eta\sigma\beta}Q_{\rho\sigma\alpha\beta}.              \nonumber
\end{eqnarray} 
The expressions for the Ricci tensor are given in Appendix 2.

\initiate
\section{Geometry of the truncated Heisenberg space}

From the discussion of the previous sections we see that geometry is defined by the choice 
of $p_\alpha$: Therefore at the first sight it appears to be rather arbitrary. However we 
have also seen that the additional requirements like metric compatibility or vanishing of 
the torsion induce additional constraints, for example $F^{(\alpha\beta\gamma)} =0 $. 
We will in addition impose the condition that in the limit $n\to\infty$
both algebra and differentials tend to the values which they have in the Heisenberg 
algebra. This fixes the momenta almost uniquely,
\begin{equation}
 \epsilon p_1 =i \mu^2 y, \qquad \epsilon p_2 =-i\mu^2 x,\qquad \epsilon p_3 =i\mu 
(\mu z- \frac 12) .
\end{equation} 
The momentum algebra is therefore given by
\begin{eqnarray}
 && [p_1,p_2] =\frac {\mu^2}{2i\epsilon} + \mu p_3,              \label{ppp}\\[4pt]
&& [p_2,p_3] = \mu p_1 -i\epsilon (p_1p_3+p_3p_1), \nonumber \\[4pt]
&& [p_3,p_1] = \mu p_2  -i\epsilon (p_2p_3 +p_3p_2) ,\nonumber
\end{eqnarray}
while the nonvanishing structure coefficients have the values
\begin{equation}
 K_{12} =\frac {\mu^2}{2},\qquad F^1{}_{23} =\mu,\qquad
 Q^{13}{}_{23} = \frac 12, \qquad  Q^{23}{}_{31} = \frac 12    ,     \label{struc}
\end{equation}
and those obtained by symmetries, for example $Q^{23}{}_{31} = Q^{32}{}_{31}=- Q^{23}{}_{13} $.
We shall assume that the truncated Heisenberg space has diagonal metric of Euclidean 
signature, $(+++)$ and we will take the connection in the form (\ref{connection}). 
From (\ref{T}) we obtain that the nonvanishing  $T_{\alpha\beta\gamma\delta}$ are
\begin{equation}
\begin{array}{lll}
 T_{1332} = 2,&T_{1233}=2 ,& T_{2133} =-2,\\[6pt]
 T_{2331}= -2 ,&
T_{3132}=2,&T_{3231}= -2 ,
\end{array} \nonumber
\end{equation} 
so the connection 1-form is given by
\begin{eqnarray}
 && \omega_{12} = -\omega_{21} = (-\frac \mu 2 + 2i\epsilon p_3) \theta^3 = \mu\, 
(\frac 12-2\mu z)\theta^3 ,               \label{Con}\\
&& \omega_{13} = -\omega_{31} =\frac \mu 2 \,\theta^2 +2i\epsilon  p_2\theta^3 = 
\frac \mu 2\, \theta^2 +2 \mu^2 x\theta^3 ,\nonumber \\
&& \omega_{23} = -\omega_{32} = -\frac \mu 2 \,\theta^1 - 2i \epsilon p_1\theta^3
= -\frac \mu 2 \, \theta^1 +2\mu^2 y\theta^3 .\nonumber
\end{eqnarray} 
The scalar curvature, from  (\ref{R}), is
\begin{equation}
R = R_0 = \frac{11\mu^2 }{2} +4 i \epsilon \mu p_3 - 8 (i\epsilon)^2( p_1^2 +p_2^2 )
=\frac{15\mu^2}{2} - 4 \mu^3 z - 8\mu^4  (x^2 +y^2)  .         \label{scalar}
\end{equation}

It is important to understand exactly the properties of the embedding of the 
hyperplane $z=0$ into the truncated Heisenberg algebra. From the 
point of view of the algebra it is a two-dimensional noncommutative space 
generated by $x$ and $y$ or by $p_1$ and $p_2$. As on the subspace $\ p_3 = -\frac{i\mu}{2\epsilon}\, $ 
we also have for all functions  $\ e_3 f =[p_3,f]=0 $, consistently with 
(\ref{ppp}). However, the  contangent space is three-dimensional. Though from
\begin{eqnarray}
&&dx =(1-\mu z)\theta^1+\mu^2 (yz+zy)\theta^3 ,\label{emu} \\[4pt]
&& dy=(1-\mu z)\theta^2 - \mu^2(xz+zx)\theta^3 ,\nonumber \\[4pt]
&& dz= \mu^2 (xz+zx)\theta^1 + \mu^2 (yz+zy)\theta^2 \nonumber
\end{eqnarray}
for $z= 0$ we obtain $dx =\theta^1$, $dy =\theta^2$, $dz =0$, it is clear that it is
impossible to express one of the basic 1-forms $\theta^\alpha$ in terms of the other 
two, in order to replace it for example in (\ref{Con}) or in some other formula.

The fact that the space of 1-forms can have different (higher) dimensionsionality 
from the space of coordinates is known in noncommutative geometry. One typical example 
is the fuzzy sphere \cite{fuzzy}, where similarly the space `itself' 
has dimension two whereas the cotangent space is of dimension three. Though for 
the fuzzy sphere this difference does not show in the  calculation
of the scalar curvature, it does have important consequences on the construction of 
 gauge theory, \cite{Grosse:1992bm}. 
The gauge potential on the fuzzy sphere has the natural dimension three, and all three 
degrees of freedom are needed  to establish the relation with the dynamics 
of branes on $S^3$ in string theory for example, \cite{Alekseev:2000fd}. To discuss the commutative limit on the other hand, one has to impose an additional 
constraint, \cite{fixing}. In our truncated Heisenberg case we have a 
three-dimensional cotangent space too.

The value of the connection on  $z=0$ is given by
\begin{eqnarray}
 &&\omega_{12} =-\omega_{21} =\frac \mu 2 \, \theta^3, \label{omegalim}\\
&&\omega_{13} = -\omega_{31} = \frac \mu 2 \, \theta^2    + 2\mu^2x\theta^3, \nonumber \\
&& \omega_{23} = -\omega_{32} =-\frac \mu 2 \, \theta^1 + 2\mu^2 y\theta^3 ,  \nonumber
\end{eqnarray} 
and the corresponding value of the scalar curvature is
\begin{equation}
R = \frac{15 \mu^2}{2}  - 8 \mu^4 (x^2 +y^2) .                  \label{a,b}
\end{equation}

\initiate
\section{Relation to the Grosse-Wulkenhaar action}

Having the value of the scalar curvature (\ref{a,b}) it is not difficult to recognize 
the relation between the Grosse-Wulkenhaar action (\ref{GW4})
and the action for the scalar field on a curved space.
 In the notation of \cite{Grosse:2004yu},
$\tilde x_\mu  = ip_\mu  $ so  we have $\tilde x^\mu \tilde x_\mu =-\frac{\mu^4}{\epsilon^2}x^\mu  x_\mu$.  Using  the cyclicity under the integral we have
\begin{equation}
 \int{\tilde x}^\mu\varphi \, {\tilde x}_\mu\varphi = \int -\frac 12 [p_\mu, \varphi][ p^\mu ,\varphi] + {\tilde x}^\mu {\tilde x}_\mu \varphi^2 ,
\end{equation} 
and the Grosse-Wulkenhaar action can be rewritten as
\begin{equation}
 S=  \int \frac 12 \,(1-\frac {\Omega^2}{ 2})\p_\mu \varphi \,\p^\mu \varphi +\frac {m^2}{2 }\varphi^2  +\frac{\Omega^2}{2} {\tilde x}^\mu {\tilde x}_\mu\varphi \varphi +\frac{\lambda}{4!}\varphi^4 .
                                                \label{gw1}
\end{equation}
On the other hand, the action for the scalar field  non-minimally coupled to the curvature 
is given by
\begin{equation}
 S^\prime = \int \sqrt g \left( \frac 12\,\p_\mu \varphi \,\p^\mu \varphi +\frac {M^2}{2 }\varphi^2  - \frac \xi 2 R\varphi^2 +\frac{\Lambda}{4!}\varphi^4  \right) .                \label{curved}
\end{equation}
We have seen already that for $z=0$, $\, e_\alpha = \delta^\mu_\alpha \p_\mu\,$ for $\mu =1,2$, $\,e_3 =0\,$ and $\,\sqrt g =1$. Therefore we find that (\ref{gw1}) and (\ref{curved}) are the same 
up to an overall rescaling
\begin{equation}
 S=\kappa S^\prime ,
\end{equation} 
if we identify
\begin{equation}
  1-\frac{\Omega^2}{2} =  \kappa, \qquad m^2 =\kappa(M^2-\xi a) ,\qquad \frac{\Omega^2\mu^4}{\epsilon^2} =\kappa\xi b, \qquad
\lambda =\kappa\Lambda     ,                 \label{ren}
\end{equation} 
and  $a$ and $b$ from (\ref{a,b}),  $\ a=\frac{15\mu^2}{2}$, $\, b=8\mu^4$.

The constant part of the curvature renormalizes the mass of the scalar field, 
while the space-dependent part gives the harmonic oscillator potential.
The coupling constant $\xi$  is not a priori fixed but can be related to $\Omega$. 
If we identify the two actions at the self-duality point $\Omega =1$ we obtain
\begin{equation}
 \xi = \frac {\Omega^2\mu^4}{\epsilon^2\kappa b} = \frac{1}{4\epsilon^2} .
\end{equation} 
Note that if $M=0$ in the initial action (\ref{curved}) our mechanism induces a mass term with negative sign and thus we obtain a Lagrangian containing the Higgs potential and spontaneous symmetry breaking.

It is not difficult to generalize the given construction from two to four spatial dimensions 
and reach the same conclusion for the four-dimensional Grosse-Wulkenhaar model. 
The key element is that there exist a set of coordinates in which noncommutativity $J^{\mu\nu}$ has the canonical, block-diagonal form
\begin{equation}
  J^{\mu\nu} 
\sim
 \begin{pmatrix}
  &1 & & \cr
-1 &  & & \cr
& &  & 1\cr 
& & -1 & 
\end{pmatrix} .
\end{equation} 
In these coordinates the four-dimensional noncommutative space $\c{A}$ is a direct product 
of two two-dimensional spaces, $\c{A} =\c{A}^{(1)}\otimes \c{A}^{(2)}$; $\c{A}^{(1)}$ and $ \c{A}^{(2)}$ commute. Then, finite-matrix approximations to $\c{A}$ can be defined by taking direct products of the approximations to $\c{A}^{(1)}$ and $ \c{A}^{(2)}$ which are described above. The four-dimensional $\c{A}$ is a subspace of the six-dimensional space defined by $ z^{(1)}=0,\ z^{(2)}=0$. Clearly, as the product spaces commute, the scalar curvature  is the sum of 
curvatures
\begin{equation}
 R = R^{(1)} + R^{(2)} ;
\end{equation} 
similar holds true for the Laplacian. Thus one can make the same identification of the constants (\ref{ren}) and of the actions in four dimensions as one does in two.

The procedure to obtain the four-dimensional action described above is the simplest one can think of. It apparently breaks the symmetry among $x^\mu$: but in fact this symmetry is broken from the start by the values of the components of $J^{\mu\nu}$. It would be interesting to find another,  minimal in the sense of dimensionality and finite, approximation to the four-dimensional Heisenberg algebra.

\section{Concluding remarks}

To summarize: We have shown that it is possible, through a sequence of 
matrix representations, to define a noncommutative space which has the same algebra
but a different geometry from the Moyal-deformed space. Specific properties
of the described space allow the interpretation of the oscillator term in the 
Grosse-Wulkenhaar model as a coupling to the background curvature.
In particular, the given picture explains the absence of the 
translation invariance in the Grosse-Wulkenhaar action: The underlying curved 
space does not possess it. The performed construction can be
extended  easily to any even-dimensional space.

An important technical detail was the difference in dimensionality
between the basic space and the cotangent space.
This possibility, typical for noncommutative geometry, 
has been studied before for the fuzzy sphere, \cite{fuzzy,Grosse:1992bm,
Alekseev:2000fd, fixing}. Its characteristic consequences appear whenever 
one deals with fields which `live' in the cotangent space, 
for example  gauge fields or  linear connection, and are in some ways similar
to the Kaluza-Klein reduction. Another fact 
important to stress is that the set of intermediate matrix spaces discussed 
above was not introduced arbitrarily; rather, the matrix base was used to establish the first 
proof of renormalizability of the model \cite{Grosse:2003nw}, relying on estimating the decay
properties of the propagator.
%these matrices were used 
%to estimate the behavior of the propagator in the first proof of renormalizability 
%of the model \cite{Grosse:2003nw}. 

An interesting point is that we have identified a model
in which the field theory renormalization is (or can be interpreted to be) 
done effectively by the curvature. The old idea of 
Pauli, Deser and others \cite{GR} that gravity can regularize field theory is 
here realized in a very specific way, in the setting of noncommutative geometry. 
However it does not directly correspond to the common intuition that regularization 
works through the uncertainty relations; the regularization is rather indirect, 
through the curvature, \cite{Madore:1996bb}. There is of course another 
ingredient of the given construction which is hard to disentangle from its geometric 
aspects: the finiteness of the repesentation. This element might be 
even primary in considering renormalizability, and perhaps indicates an 
advantage of the theories which can be regularized through matrix models.

The model described in the paper opens, in our opinion, interesting 
new possibilities to understand relations between noncommutative gravity and 
noncommutative field theory. One possibility to interpret the oscillator
term, given previously in \cite{LS}, is to relate it to the coupling
of the complex scalar to the external magnetic field.
Here it is the external gravitational field which couples 
to the scalar $\varphi$. (In our approach both real and complex scalar fields
have the same behavior, as their coupling to gravity is of the same form.) 
If the geometric interpretation
 has a deeper physical meaning, it should provide also a description for 
the other fields, for example gauge fields or spinors.  
Given the fixed background geometry the corresponding actions
 should be straightforward to define;
we will analyze properties of such models in our future work. Another 
important aspect which should be addressed in the future is to understand 
what is exactly the 
role of the of the Langmann-Szabo duality in the given framework and 
in particular, whether there is a relation to the frame formalism.

{\bf Acknowledgment}\ \ The authors would like to thank J. Madore and 
H. Grosse for  exciting discussions during the work on this manuscript. 
This  work of was supported in part by ESF grant 2747 through the Quantum 
Gravity Network and by 141036 grant of MNTR, Serbia.
%\bibliographystyle{../latex-styles/utphys}
%\bibliography{../tuw}

\section*{Appendix 1}

We mentioned briefly in Section 3 that the relations (\ref{**}) need not be included in the
algebra of  coordinates (\ref{zz}), or of the momenta (\ref{ppp}). The reason  
 is, that these relations are not expressed in the form of commutators -- while
the differential calculus is,
$df =[p_\alpha,f] \theta^\alpha$. 
On the other hand, (\ref{**}) are consistent that is stable under 
differentiation, for example
$d(Pa) = 0$, $d(P^2-P) =0$ etc.; this can be checked easily.

However the projector condition from (\ref{**}) can  be used to modify
the algebra (\ref{zz}) and to write it  in another, also quadratic, form.
(The quadratic form is preferred because then the identification of  momenta
is much easier.)
Using $\ n\mu z = (\mu z)^2\ $ we can rewrite (\ref{zz}) as
\begin{eqnarray}
 &&[x,y] = i \epsilon \mu^{-2}(1-\frac{\mu ^2 z^2 }{n} ) ,                                            \label{zzprime}\\[4pt]
&& [x,z]  = i\epsilon(yz+zy) ,                                      \nonumber\\[4pt]
&& [y,z] = - i\epsilon(xz+zx)            .                           \nonumber
\end{eqnarray} 
The linear terms in (\ref{zzprime}) are absent, so it is simpler and perhaps
more natural to choose the momenta as:
\begin{equation}
 \epsilon p_1^\prime =i \mu^2 y, \qquad \epsilon p_2^\prime =-i\mu^2 x,\qquad \epsilon p_3^\prime =i\mu ^2 z .                         \label{momprime}
\end{equation} 
Of course, this changes the differential calculus but not much as we shall
shortly see. We are interested in the curvature. The momentum algebra is now given by
\begin{eqnarray}
 && [p_1^\prime,p_2^\prime] =\frac {\mu^2}{i\epsilon} -\frac{i\epsilon}{n}( p_3^\prime)^2,              \label{ppprime}\\[4pt]
&& [p_2^\prime,p_3^\prime] =  -i\epsilon (p_1^\prime p_3^\prime +p_3^\prime p_1^\prime), \nonumber \\[4pt]
&& [p_3^\prime,p_1^\prime] =  -i\epsilon (p_2^\prime p_3^\prime +p_3^\prime p_2^\prime) ,\nonumber
\end{eqnarray}
and the corresponding nonvanishing structure coefficients are 
\begin{equation}
 K^\prime_{12} =\mu^2,\qquad
 Q^{\prime 13}{}_{23} = \frac 12, \qquad  Q^{\prime 23}{}_{31} = \frac 12    ,   
 \qquad  Q^{\prime 33}{}_{12} = \frac {1}{2 n} .   \label{strucprime}
\end{equation}
Calculating the scalar curvature from the connection defined in the same manner
as before, we obtain
\begin{eqnarray}
&& R^\prime  = 8\mu^2 - 8 (i\epsilon)^2\left( (p_1^\prime)^2 +(p_2^\prime)^2 \right)
+\frac{4(i\epsilon)^3}{n}\,[p_1^\prime,p_2^\prime]           \label{scalarprime}
\\[4pt]  &&  \phantom{R^\prime } =
 8\mu^2  - 8\mu^4  (x^2 +y^2)  -\frac{4\epsilon^2}{n}\mu^2 + \frac{4\epsilon^2}{n^2} \mu^4 z^2 .
\nonumber
\end{eqnarray} 
On the subspace $z=0$, $n\to \infty$ it reduces to
\begin{equation}
R^\prime = 8\mu^2  - 8\mu^4  (x^2 +y^2)  .      
\end{equation} 

The result is rather interesting:  the desired
quadratic dependence on coordinates appears again, though the value of the 
scalar curvature is not exactly the same. This change one  can  attribute
to the change of the momenta (\ref{momprime}), that is to the change of the differential
$d$.  It might  be interesting to compare in some detail the respective
connections $\omega$ and $\omega^\prime$. Invariance and 
properties of the geometric characteristics of noncommutative spaces under 
the change of generators of the algebra are certainly an important topic which deserves 
further study.

\section*{Appendix 2}

The components of
the Ricci curvature which correspond to the Riemann tensor (\ref{rr}) are given by 
\begin{eqnarray}
 && 2 {R_0}_{\beta\sigma}  = 2 {R_0}^\alpha{}_{\beta\alpha\sigma} = T^{\alpha\gamma}{}_{\sigma\beta} K_{\alpha\gamma} - T^{\alpha\gamma}{}_{\alpha\beta} K_{\sigma\gamma} -\frac 14 F^\alpha{}_{\gamma\beta} F^\gamma{}_{\alpha\sigma}
\nonumber \\[6pt] 
&& \phantom{2R}
+\ep p_\zeta \Big( F^\zeta{}_{\delta\gamma} T^{\alpha\gamma}{}_{\sigma\beta} -F^\zeta{}_{\sigma\gamma} T^{\alpha\gamma}{}_{\alpha\beta} -F^\gamma{}_{\alpha\sigma} T^{\alpha\zeta}{}_{\gamma\beta} 
- \frac 12 F^\alpha{}_{\sigma\gamma} T^{\gamma\zeta}{}_{\alpha\beta}
\nonumber \\[6pt] 
&& \phantom{2R +\ep p_\zeta \Big(}
-\frac 12 F^\alpha{}_{\gamma\beta} T^{\gamma\zeta}{}_{\alpha\sigma} 
+\frac 12 F^\alpha{}_{\gamma\beta} T^{\gamma\zeta}{}_{\sigma\alpha}  -
\frac 12 F^\gamma{}_{\alpha\beta} T^{\alpha\zeta}{}_{\sigma\gamma} + \frac 12 F^\gamma{}_{\sigma\beta} T^{\alpha\zeta}{}_{\alpha\gamma } \Big)
\nonumber \\[6pt] 
&& \phantom{2 R}
+ (\ep)^2 p_\zeta p_\eta\Big( 
-T^{\alpha\zeta}{}_{\gamma\beta} T^{\gamma\eta}{}_{\alpha\sigma} +T^{\alpha\zeta}{}_{\gamma\beta} T^{\gamma\eta}{}_{\sigma\alpha} +T^{\alpha\zeta}{}_{\alpha\gamma} T^{\gamma\eta}{}_{\sigma\beta} -T^{\alpha\zeta}{}_{\sigma\gamma} T^{\gamma\eta}{}_{\alpha\beta}
\nonumber \\[6pt] 
&& \phantom{2 R+ (\ep)p_b\Big( }
+ \frac 12 T^{\alpha\gamma}{}_{\sigma\beta} T^{\zeta\eta}{}_{\alpha\gamma} - \frac 12 T^{\alpha\gamma}{}_{\sigma\beta} T^{\zeta\eta}{}_{\gamma\alpha} -\frac 12 T^{\alpha\gamma}{}_{\alpha\beta} T^{\zeta\eta}{}_{\sigma\gamma} + \frac 12 T^{\alpha\gamma}{}_{\alpha\beta} T^{\zeta\eta}{}_{\gamma\sigma}  \Big) , \nonumber
\end{eqnarray}
\begin{eqnarray}
 &&{R_1}_{\beta\nu}=  \Big( T^{\alpha\gamma}{}_{\sigma\beta} K_{\rho\gamma} + \frac 14 F^\alpha{}_{\rho\gamma} F^\gamma{}_{\sigma\beta}
\qquad\qquad\qquad \qquad\qquad\qquad \qquad \qquad\qquad\qquad \nonumber       \\[6pt]
&&\phantom{2R}+\ep p_\zeta( F^\zeta{}_{\rho\gamma}T^{\alpha\gamma}{}_{\sigma\beta} +\frac 12 F^\alpha{}_{\rho\gamma}T^{\gamma\zeta}{}_{\sigma\beta} +\frac 12 F^\gamma{}_{\sigma\beta}T^{\alpha\zeta}{}_{\rho\gamma}
)  \nonumber \\[6pt]
&&\phantom{2R}+(\ep)^2 p_\zeta p_\eta (-2T^{\alpha\gamma}{}_{\sigma\beta}Q^{\zeta\eta}{}_{\rho\gamma}+T^{\alpha\zeta}{}_{\rho\gamma}T^{\gamma\eta}{}_{\sigma\beta}
)\Big) Q^{\rho\sigma}{}_{\alpha\nu} .\nonumber
\end{eqnarray} 
In principle, they enter the field actions when gravity couples to the other fields. On the subspace
$z=0$ of the truncated Heisenberg space the value of the Ricci tensor is
\begin{equation} 
R_{\alpha\beta} = \left(\begin{array}{ccc}
\frac{3\mu^2}{2}-4\mu^4 x^2 
& -2\mu^4(xy+yx)+i\frac{\epsilon\mu^2}{4}
& 2\mu^3 y +2i\epsilon \mu^3 x 
\\[10pt]
- 2\mu^4(xy+yx)-i\frac{\epsilon\mu^2}{4}
& \frac{3\mu^2}{2}-4\mu^4 y^2 
& -2\mu^3 x +2i\epsilon \mu^3 y 
\\[10pt]
2\mu^3 y - 2i\epsilon \mu^3 x & -2\mu^3 x - 2i\epsilon \mu^3 y &
\frac{9\mu^2}{2} -4\mu^4(x^2+y^2) 
 \end{array}\right) . \nonumber
\end{equation}

\end{document}